\documentclass[pra,twocolumn,amsmath,amssymb, superscriptaddress]{revtex4}

\usepackage{graphicx}
\usepackage{dcolumn}
\usepackage{bm,color}

\newcommand{\half}{{\textstyle \frac{1}{2}}}

\begin{document}

\title{A quantum stochastic calculus approach to modeling double-pass atom-field coupling}

\author{Gopal Sarma} \email{gsarma@stanford.edu}
\author{Andrew Silberfarb}
\author{Hideo Mabuchi}
\affiliation{Physical Measurement and Control, Edward L. Ginzton Laboratory, Stanford University, Stanford, CA 94305}

\date{\today}

\begin{abstract}
We examine a proposal by Sherson and M\o lmer to generate polarization-squeezed light in terms of the quantum stochastic calculus (QSC).  We investigate the statistics of the output field and confirm their results using the QSC formalism.  In addition, we study the atomic dynamics of the system and find that this setup can produce up to 3 dB of atomic spin squeezing.
\end{abstract}

\maketitle

\section{Introduction}
Many recent experiments in quantum information protocols and precision metrology have utilized the optical Faraday rotation of light passing through spin polarized atomic samples.  Of particular interest to our discussion is a recent article by Sherson and M\o lmer, in which it is shown that polarization squeezed light can be generated by sending a cw beam or pulse of linearly polarized light through an atomic gas twice in different directions (See Fig. 1) \cite{ShM06}.  For a proposal involving a similar experimental setup in the context of atomic magnetometry, see \cite{JM07}.   

The aim of this paper is to analyze this experiment using the quantum stochastic calculus (QSC).  After a brief introduction to QSC, we describe the system in terms of a quantum stochastic differential equation (QSDE).  Using this model, we derive the atomic and field dynamics of the double-pass system, confirming their results, and furthermore, showing that the resultant atomic states are spin-squeezed up to 3 dB.  

\begin{figure}[t]
\includegraphics[width=0.3\textwidth]{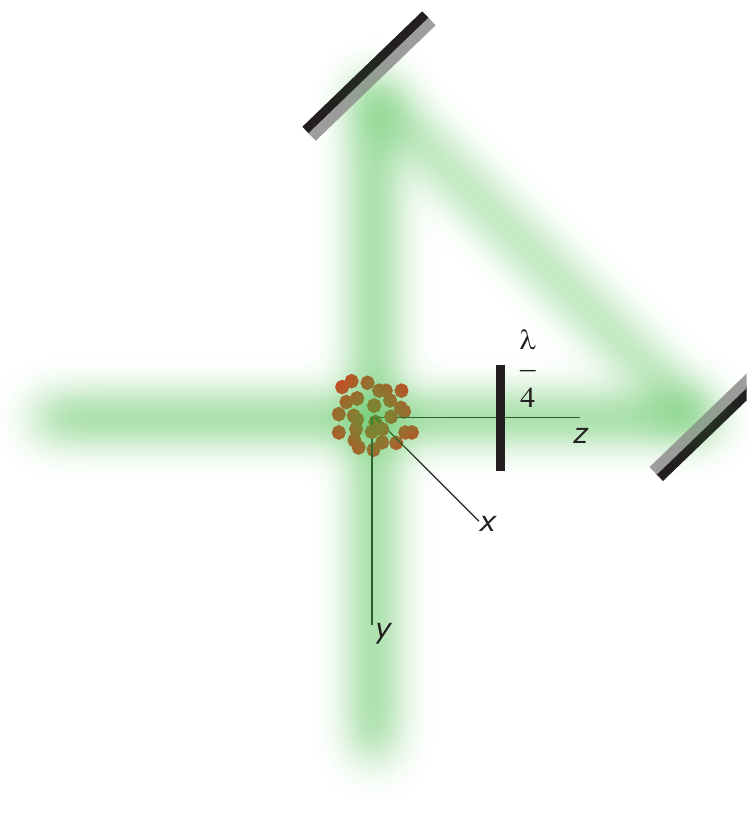}
\caption{\small
  Sherson and M\o lmer's setup for the generation of squeezed light. 
  After transmission through a gas of spin polarized atoms, the polarization 
  of the cw beam or pulse of linearly polarized light is rotated with a quarter wave plate, and it is
  then transmitted through the gas a second time from a different
  direction.} 
 \label{fig:setup}
\end{figure}

\section{The Model}\label{sec model}

We consider an atomic gas in interaction with 
the quantized electromagnetic field.  (Our development of the model closely follows the exposition in \cite{Bas06}, where a single qubit, rather than an atomic gas, is studied).  As the interaction between the gas and the field is symmetric, we can characterize the atomic ground state by the collective spin vector $\hat{\vec{J}}$.  Furthermore, assuming that the atoms are strongly spin polarized in the $\hat{x}$ direction, we can apply the Holstein-Primakoff approximation and introduce the dimensionless variables $(x_{at}, p_{at}) = (\hat{J_{y}}, \hat{J_{z}}) / \sqrt{J_{x}}$.  Setting $\hbar = 1$, these obey the canonical commutation relationship $[x_{at}, p_{at}] = i$. 
 
The atomic gas is described by the Hilbert space of quadratically integrable 
functions $L^{2}({\mathbb{R}})$ and the electromagnetic field by the \emph{symmetric Fock space} $\mathcal{F}$ over $L^2(\mathbb{R})$ (Hilbert space of single photon wave functions), i.e.\
  \begin{equation*}
  \mathcal{F} := \mathbb{C} \oplus \bigoplus_{k = 1}^\infty L^2(\mathbb{R})^{\otimes_s k}.
  \end{equation*}
The Fock space $\mathcal{F}$ allows us to characterize superpositions of field-states 
with different numbers of photons. The joint 
system of the gas and field together is given by the 
Hilbert space $L^{2}({\mathbb{R}})\otimes\mathcal{F}$.

We examine the interaction between the gas and the electromagnetic field in the weak coupling limit \cite{Gou04}, \cite{Gou05}, \cite{AFLu90} so that in the interaction picture, the unitary dynamics of 
the gas and the field together is given by a quantum 
stochastic differential equation (QSDE) as described by Hudson 
and Parthasarathy \cite{HuP84}.   
 \begin{equation}\label{eq QSDE}
dU_{t} = \{-L^{*}dA_{t} + LdA_{t}^{*} - \half L^{*}Ldt -iHdt\}U_{t},
 \end{equation}    
 with $L = \frac{\alpha(p - ix)}{\sqrt{2}}$, $H = \textstyle{\frac{1}{4}} \alpha^{2}(px + xp)$, and $U_{0} = I$. 
(For a derivation of the double-pass QSDE, see Appendix \ref{doublepass}).  Here, the operators $L$ and $L^{*}$ are proportional to the the annihilation and creation operators on the atomic gas system and $A_t$ and $A^*_t$ denote the field annihilation and creation processes. 
Note that the evolution
$U_{t}$ acts nontrivially on the combined system $L^{2}(\mathbb{R})\otimes\mathcal{F}$. 
That is, $L$ and $A_{t}$ are understood to designate
the single system-operators $L\otimes I$ and $I\otimes A_{t}$ respectively.  
Equation \eqref{eq QSDE} 
should be taken as a shorthand for the integral equation
  \begin{equation*}
  \begin{split}
  U_{t} = I& - \int_0^tL^{*} U_{\tau} dA_{\tau} 
   + \int_0^tLU_{\tau}dA^{*}_{\tau} 
   - \half \int_0^tL^{*}L U_{\tau}d\tau \\
   &-i \int_0^tH U_{\tau}d\tau, 
   \end{split}
  \end{equation*} 
where the integrals on the right-hand side are Hudson-Parthasarathy
stochastic integrals, and where the stochastic increments satisfy the following rule \cite{HuP84}, \cite{Par92}: 

\label{Itorule}\textbf{Quantum It\^o rule}
Let $X_t$ and $Y_t$ be stochastic integrals of the form 
\begin{equation*}\begin{split}
&dX_t = C_t dA_t + D_t dA_t^* + E_tdt \\
&dY_t = F_t dA_t + G_t dA_t^* + H_tdt
\end{split}\end{equation*}
for stochastically integrable processes $C_t, D_t, E_t, F_t, 
G_t$ and $H_t$ (see \cite{HuP84}, \cite{Par92} for definitions). 
Then the process $X_tY_t$ is itself a stochastic integral and satisfies
  \begin{equation*}
  d(X_tY_t) = X_tdY_t + (dX_t)Y_t + dX_tdY_t,
  \end{equation*}
where $dX_tdY_t$ should be evaluated according to 
the quantum It\^o table:
\begin{center}
{\large \begin{tabular} {l|lll}
 & $dA_t$  & $dA^*_t$ & $dt$\\
\hline 
$dA_t$ & $0$ & $dt$ & $0$\\
$dA^*_t$ & $0$  & $0$ & $0$\\
$dt$ & $0$ & $0$ & $0$
\end{tabular} }
\end{center}
i.e.\ $dX_tdY_t = C_tG_tdt$. 

We now describe a simple notation (introduced in \cite{Bas06}) for expressing differentials of products of stochastic integrals:
Let $\{Z_i\}_{i=1,\dots, p}$ be stochastic integrals. Then we write 
 \begin{equation*} 
 d (Z_1Z_2\dots Z_p)= \sum_{\substack{\nu\subset\{1,\dots, p\} \\ \nu \neq \emptyset}}\{\nu\},
 \end{equation*}
where the sum is taken over non-empty subsets of $\{1,\dots, p\}$. For 
$\nu=\{i_1,\dots, i_k\}$, the term $\{\nu\}$ is obtained by differentiating only the terms with indices 
in the set $\{i_1,\dots, i_k\}$ and preserving the order of the factors in the product. 
For instance, consider the differential $d(Z_1Z_2Z_3)$, which contains terms of type
$\{1\}$, $\{2\}$, $\{3\}$, $\{12\}$, $\{13\}$, $\{23\}$ and $\{123\}$. Applying the It\^o rule, we have
$\{2\}=Z_1(dZ_2)Z_3$, $\{13\}=(dZ_1)Z_2(dZ_3)$, $\{123\}=(dZ_1)(dZ_2)(dZ_3)$,
and so on. (To avoid confusion with the notation for references, we have used curly braces \{\} instead of the square braces [ ] used in \cite{Bas06}).  

\section{Derivation of the Joint Characteristic Functions}\label{sec joint}
In this section, we derive the joint characteristic functions for the combined atom-field system.  Although we will not be using the input-output formalism \cite{GC84}, we present it here for comparison with \cite{ShM06}:
\begin{align*}
x_{at}^{in}&= \hat{J_{y}} / \sqrt{J_{x}}& 
dx_{ph}^{in}&= \frac{dA_{t} + dA_{t}^{*}}{\sqrt{2}}& \\
x_{at}^{out}&= U_{t}^{*}x_{at}^{in}U_{t}& 
dx_{ph}^{out}&= \frac{d\left[U_{t}^{*}(A_{t} + A_{t}^{*})U_{t}\right]}{\sqrt{2}}& \\
\end{align*}
\begin{align*}
p_{at}^{in}&= \hat{J_{z}} / \sqrt{J_{x}}& 
dp_{ph}^{in}&= \frac{dA_{t} - dA_{t}^{*}}{i\sqrt{2}}& \\
p_{at}^{out}&= U_{t}^{*}p_{at}^{in}U_{t}&
dp_{ph}^{out}&= \frac{d\left[U_{t}^{*}(A_{t} - A_{t}^{*})U_{t}\right]}{i\sqrt{2}}&
\end{align*}
For the sake of simplicity in later computations, we have defined $x_{ph}^{out}$ and $p_{ph}^{out}$ in such a way that $[x_{ph}^{out}, p_{ph}^{out}] = it$.  As will be apparent later, once we have calculated the variances for these operators, we rescale by a factor of $1/t$ to obtain the actual variances for the modes that we are interested in (i.e. the squeezed and anti-squeezed modes).  Applying the It\^ o rule, and evaluating the products of stochastic integrals using the It\^ o table, we can express the above input-output relations in the language of \cite{ShM06}:
\begin{eqnarray}
x_{ph}^{out}(t) &=&x_{ph}^{in}(t) + \alpha p_{at}^{out}(t) \\
p_{ph}^{out}(t) &=& p_{ph}^{in}(t) - \alpha x_{at}^{out}(t) \\
\frac{dx_{at}^{out}(t)}{dt} &=& \alpha  p_{ph}^{in}(t) \\
\frac{dp_{at}^{out}(t)}{dt} &=& -\alpha(x_{ph}^{in}(t) + \alpha p_{at}^{out}(t))
\end{eqnarray}
As an example, we derive the expression for $x_{ph}^{out}(t)$:  
\begin{eqnarray*}
x_{ph}^{out}dt &=& \frac{d\left[U_{t}^{*}(A_{t} + A_{t}^{*})U_{t}\right]}{\sqrt{2}} \\
&=& \{1\} + \{2\} + \{3\} + \{12\} + \{23\} + \{13\} + \{123\} 
\end{eqnarray*}
A simple calculation shows that the terms \{1\}, \{3\}, and \{13\} sum to zero, and from the It\^ o table, we see that third powers of increments (i.e. \{123\}) vanish, leaving us with \{2\}, \{12\} and \{23\} to calculate:
\begin{eqnarray*}
\{2\} &=& \frac{dA_{t} + dA_{t}^{*}}{\sqrt{2}} \\
\{12\} &=& dU_{t}^{*}\frac{dA_{t} + dA_{t}^{*}}{\sqrt{2}}U_{t} =  U_{t}^{*}\frac{L^{*}}{\sqrt{2}}U_{t}dt \\
\{23\} &=& \{12\}^{*} = U_{t}^{*}\frac{L}{\sqrt{2}}U_{t}dt
\end{eqnarray*}
Summing the preceding terms, and substituting for $x_{ph}^{in}$, we obtain:
\begin{eqnarray*}
\{2\} + \{12\} + \{23\} &=& \frac{dA_{t} + dA_{t}^{*}}{\sqrt{2}} + U_{t}^{*}\frac{(L^{*} + L)}{\sqrt{2}}U_{t}dt \\
&=&x_{ph}^{in}dt + U_{t}^{*}\alpha p_{at}^{in}U_{t}dt \\
\Rightarrow x_{ph}^{out}(t) &=&x_{ph}^{in}(t) + \alpha p_{at}^{out}(t)
\end{eqnarray*}
For comparison with the results in \cite{ShM06}, $\alpha$ should be equated with $\kappa$, and as we have ignored damping, $\tau$ should be set to 1. 

Define $F[t,k,l]$ and $G[t,k,l]$ as follows (see, for instance, \cite{Bar85}): 
\begin{eqnarray}
F[t,k,l]:=\langle v \otimes \Phi \mid U_{t}^{*}(e^{ilp}\otimes e^{ik\frac{(A_{t} + A_{t}^{*})}{\sqrt{2}}})U_{t} \mid v \otimes \Phi \rangle& \\
G[t,k,l]:=\langle v \otimes \Phi \mid U_{t}^{*}(e^{ilx}\otimes e^{k\frac{(A_{t} - A_{t}^{*})}{\sqrt{2}}})U_{t} \mid v \otimes \Phi \rangle&
\end{eqnarray}
In this notation, $F$ denotes the joint characteristic function for $\frac{A_{t}+A_{t}^{*}}{\sqrt{2}}$ and $p$, while $G$ denotes the joint characteristic function for $\frac{A_{t}-A_{t}^{*}}{i\sqrt{2}}$ and $x$.  While in general, we would need to calculate the joint characteristic function for all 4 variables, in the particular system we are studying, the function decouples into two independent components.  Here, $F$ and $G$ are expectation values taken with respect to an x-polarized spin state of the atoms and the vacuum state of the field, as described in the previous section.  Since we are interested in obtaining joint characteristic functions, and not individual moments, the atomic and field operators are given by complex exponentials of the respective observables. 

We can calculate $F[t,k,l]$ and $G[t,k,l]$ by solving partial differential equations given by the following lemma:

\textbf{Lemma:}
\begin{eqnarray}
\frac{\partial F}{\partial t} & = & -\textstyle{\frac{1}{4}}(\alpha l - k)^{2}F - \alpha (\alpha l - k)\frac{\partial F}{\partial l} \\
\frac{\partial G}{\partial t} & = & -\textstyle{\frac{1}{4}}(\alpha l + k)^{2}G - \alpha k\frac{\partial G}{\partial l}
\end{eqnarray}

where  $F[0,k,l] = G[0,k,l] = e^{-\frac{l^{2}}{4}}$.

\textbf{Proof:}
Let $F(Z):=\langle v \otimes \Phi \mid U_{t}^{*}(Z\otimes e^{ik\frac{(A_{t} + A_{t}^{*})}{\sqrt{2}}})U_{t} \mid v \otimes \Phi \rangle$ so that $F[e^{ilp}] = F[t,k,l]$.  Using the notation introduced previously, we have the expression $dF[e^{ilp}] = \langle v \otimes \Phi \mid \{1\} + \{2\} + \{3\} + \{12\} + \{23\} + \{13\} + \{123\} \mid v \otimes \Phi \rangle$.  Applying the It\^o rule and noting that the third powers of increments vanish, we are left with the following differentials to calculate:
\begin{eqnarray*}
\{1\} &=& dU_{t}^{*}(e^{ilp}\otimes e^{ik\frac{(A_{t} + A_{t}^{*})}{\sqrt{2}}})U_{t} \\
\{2\} &=& U_{t}^{*}\big(e^{ilp}\otimes (\textstyle{\frac{ik(dA_{t} + dA_{t}^{*})}{\sqrt{2}}}-\textstyle{\frac{1}{4}}k^{2}dt)e^{ik\frac{(A_{t} + A_{t}^{*})}{\sqrt{2}}}\big)U_{t} \\
\{3\} &=&U_{t}^{*}(e^{ilp}\otimes e^{ik\frac{(A_{t} + A_{t}^{*})}{\sqrt{2}}})dU_{t}
\end{eqnarray*}
\begin{eqnarray*}
\{12\} &=& dU_{t}^{*} \big(e^{ilp}\otimes (\textstyle{\frac{ik(dA_{t} + dA_{t}^{*})}{\sqrt{2}}}-\textstyle{\frac{1}{4}}k^{2}dt)e^{ik\frac{(A_{t} + A_{t}^{*})}{\sqrt{2}}}\big)U_{t} \\
\{23\} &=& U_{t}^{*} \big(e^{ilp}\otimes (\textstyle{\frac{ik(dA_{t} + dA_{t}^{*})}{\sqrt{2}}}-\textstyle{\frac{1}{4}}k^{2}dt)e^{ik\frac{(A_{t} + A_{t}^{*})}{\sqrt{2}}}\big)dU_{t} \\
\{13\} &=& dU_{t}^{*}(e^{ilp}\otimes e^{ik\frac{(A_{t} + A_{t}^{*})}{\sqrt{2}}})dU_{t} \\
\end{eqnarray*}
In the above expressions, $dU_{t}$ and $dU_{t}^{*}$ are given by equation ($\ref{eq QSDE}$).   The terms $dA_{t}$ and $dA_{t}^{*}$ vanish with respect to the vacuum expectation (see reference \cite{Par92}) and we find that $\{1\}, \{3\}$, and $\{13\}$ give the following: 
\begin{equation*}
\begin{split}
\langle v \otimes \Phi& \mid \{1\}+\{3\}+\{13\} \mid v \otimes \Phi \rangle \\ &= -\half F[L^{*}Le^{ilp} + iHe^{ilp}]dt \\ &-\half F[e^{ilp}L^{*}L + ie^{ilp}H]dt + F[L^{*}e^{ilp}L]dt \\
&= F[\mathcal{L}(e^{ilp})]dt,
\end{split}
\end{equation*}
where $\mathcal{L}(Z)$ is the Lindblad operator given by 
\begin{equation*}
\mathcal{L}(Z) = -\half\{L^{*}L, Z\} + i[H, Z] + L^{*}ZL  
\end{equation*}
Recalling that $L = \frac{\alpha(p - ix)}{\sqrt{2}}$, $L^{*} = \frac{\alpha(p + ix)}{\sqrt{2}}$, and $H = \textstyle{\frac{1}{4}} \alpha^{2}(px + xp)$, we can expand the Lindblad term as follows:
\begin{equation*}
\begin{split}
F[\mathcal{L}(e^{ilp})]dt &= F[-\half\{L^{*}L, e^{ilp}\} + i[H, e^{ilp}] + Le^{ilp}L^{*}]dt \\
&= -\textstyle{\frac{1}{4}}\alpha^{2}l^{2}F[e^{ilp}]dt -i\alpha^{2}lF[pe^{ilp}]dt \\
&= -\textstyle{\frac{1}{4}}\alpha^{2}l^{2}F[t,l,k]dt -\alpha^{2}l\frac{\partial F[t,l,k]}{\partial l}dt  
\end{split}
\end{equation*}
In the last step, we used the equality $F[pe^{ilp}]$ = $-i\frac{\partial F[t,l,k]}{\partial l}$.  Summing the remaining terms, we have: 
\begin{equation*}
\begin{split}
\langle v \otimes \Phi& \mid \{2\}+\{12\}+\{23\} \mid v \otimes \Phi \rangle \\ &= -\textstyle{\frac{1}{4}}k^{2}F[e^{ilp}]dt - \frac{ik}{\sqrt{2}}F[L^{*}e^{ilp} + e^{ilp}L]dt \\
&= -\textstyle{\frac{1}{4}}k^{2}F[e^{ilp}]dt - \frac{ik}{\sqrt{2}}F[\frac{\alpha(p + ix)}{\sqrt{2}}e^{ilp} + e^{ilp}\frac{\alpha(p - ix)}{\sqrt{2}}]dt \\
&= -\textstyle{\frac{1}{4}}k^{2}F[e^{ilp}]dt + \textstyle{\frac{\alpha k l}{2}}F[t,k,l]dt + \alpha k\frac{\partial F[t,l,k]}{\partial l}dt
\end{split}
\end{equation*}
Collecting like terms, we arrive at the expression stated in the lemma.  The initial condition is obtained by noting that the atoms begin in a harmonic oscillator ground state with $\sigma_{x_{at}}^{2}=\sigma_{p_{at}}^{2}=\half$.  As the Fourier transform of a Gaussian remains Gaussian, we have that $F[0,k,l] = e^{-\frac{l^{2}}{4}}$ The derivation for $G[t,k,l]$ proceeds analogously. $\square$

We then arrive at the following solutions for the joint characteristic functions:
\begin{eqnarray}
F[t,k,l] &=& e^{-\frac{1}{2}\left[\sigma_{p_{at}}^{2}l^{2} + 2\sigma_{p_{at},x_{ph}}^{2}kl + \sigma_{x_{ph}}^{2}k^{2}\right]} \\
G[t,k,l] &=& e^{-\frac{1}{2}\left[\sigma_{x_{at}}^{2}l^{2} + 2\sigma_{x_{at}, p_{ph}}^{2}kl + \sigma_{p_{ph}}^{2}k^{2}\right]}
\end{eqnarray}
where 
\begin{eqnarray}
\sigma_{p_{at}}^{2} &=& \textstyle{\frac{1}{4}}(1 + e^{-2\alpha^{2} t})\label{pat} \\
\sigma_{p_{at},x_{ph}}^{2} &=&-\textstyle{\frac{1}{4\alpha}}(1 + e^{-2\alpha^{2}t} -2e^{-t\alpha^{2}}) \\
\sigma_{x_{ph}}^{2} &=& \textstyle{\frac{1}{4\alpha^{2}}}(3 + e^{-2\alpha^{2} t} - 4e^{-\alpha^{2}t})
\end{eqnarray}
and
\begin{eqnarray}
\sigma_{x_{at}}^{2} &=& \half(1 + \alpha^{2}t)\label{xat} \\
\sigma_{x_{at}, p_{ph}}^{2} &=& -\textstyle{\frac{1}{4}} \alpha^{3}t^{2} \\
\sigma_{p_{ph}}^{2} &=& \half t + \textstyle{\frac{1}{6}}\alpha^{4}t^{3}
\end{eqnarray}
The expressions $\sigma^{2}$ denote the variances and covariances of the respective quantities.  Recalling that $[x_{ph}^{out}, p_{ph}^{out}] = it$, we define the normalized modes $\tilde{x}_{ph} = x_{ph}/\sqrt{t}, \tilde{p}_{ph} = p_{ph}/\sqrt{t}$, so that $[\tilde{x}_{ph},\tilde{p}_{ph}] = i$.  Inserting explicit t-dependence, we then obtain the following variances for the normalized modes: 
\begin{eqnarray}
\sigma_{\tilde{x}_{ph}}^{2}(t) &=& \frac{1}{t}\sigma_{x_{ph}}^{2}(t) = \textstyle{\frac{1}{4\alpha^{2}t}}(3 + e^{-2\alpha^{2} t} - 4e^{-\alpha^{2}t})\label{xph} \\
\sigma_{\tilde{p}_{ph}}^{2}(t) &=& \frac{1}{t}\sigma_{p_{ph}}^{2}(t) = \half + \textstyle{\frac{1}{6}}\alpha^{4}t^{2}  \label{pph}
\end{eqnarray}
Comparing equations ($\ref{xph}$) and ($\ref{pph}$) to the ground state variances $\sigma_{\tilde{x}_{ph}}^{2}(0) = \sigma_{\tilde{p}_{ph}}^{2}(0) = \half$, we see that the double-pass setup generates an arbitrary degree of polarization squeezing, and from equations ($\ref{pat}$) and ($\ref{xat}$), at most 3 dB of atomic spin squeezing.  Note, however, that resultant states are not minimum uncertainty.  Although it is to be expected that for a linear system an initial Gaussian state remains Gaussian, the derivation of the joint characteristic function makes this fact explicit. \\  

\section{Discussion}
We have used a quantum stochastic model to investigate the dynamics of a system in which a laser beam is sent twice through an atomic gas in different directions.  Using this model, we reproduce Sherson and M\o lmer's results demonstrating polarization squeezing of the output field, and in addition, we show that this setup generates a maximum of 3 dB of atomic spin-squeezing.  The primary difference in using the quantum stochastic calculus is that the derivation operates entirely in the time domain.  While the time and frequency domain methods are equivalent in the simple linear systems that we examine in this paper, the advantage of quantum stochastic models is that they can be used to study non-linear systems as well.  

\section{Acknowledgments}
The authors would like to thank Luc Bouten for insightful discussions.  G.S. is supported by the ARO under Grant No. W911NF-06-1-0378. A.S. and H.M. are supported by the ONR under Grant No. N00014-05-1-0420.

\appendix
\section{Derivation of the double-pass QSDE}\label{doublepass}
We begin with the familiar single-pass QSDE's given by the Faraday interaction Hamiltonian (see for ex. \cite{Bout07}).  
\begin{eqnarray}
dU_{t}^{1} &=& \left\{-i\alpha p\frac{dA_{t} - dA_{t}^{*}}{i\sqrt{2}} - \textstyle{\frac{1}{4}}\alpha^{2}p^{2}dt\right\}U_{t}^{1} \\
dU_{t}^{2} &=&\left\{-i\alpha x\frac{dA_{t}^{*} + dA_{t}}{\sqrt{2}} - \textstyle{\frac{1}{4}}\alpha^{2}x^{2}dt\right\}U_{t}^{2}
\end{eqnarray}

To derive the QSDE for the double-pass system,  we first  divide up the light into pulses at discrete time intervals and have each ``piece'' make multiple passes.  Consider the following:
\begin{equation*}
\begin{split}
U_{t + dt}^{i} &= U_{t}^{i} + dU_{t}^{i} = M_{t}^{i}U_{t}^{i} \\
&= M_{t}^{i}M_{t-\Delta t}^{i} . . . M_{0}^{i}U_{0}^{i} = M_{t}^{i}M_{t-\Delta t}^{i} . . . M_{0}^{i}
\end{split}
\end{equation*}
where $i = 1,2$; $M_{t}^{1} = \{I - i\alpha p(\frac{dA_{t}^{*} - dA_{t}}{i\sqrt{2}}) - \textstyle{\frac{1}{4}}\alpha^{2}p^{2}dt\}$, and $M_{t}^{2} =  \{I - i\alpha x(\frac{dA_{t}^{*} + dA_{t}}{\sqrt{2}}) - \textstyle{\frac{1}{4}}\alpha^{2}x^{2}dt\}$.  The last two equalities follow by recursively applying the first and the fact that $U_{0}^{1} = U_{0}^{2} = I$.  This expression is equivalent to the series product in \cite{Yan03} and \cite{Gou07}.
We can now write the QSDE for the combined system as follows:
\begin{equation*}
\begin{split}
U_{t+dt}^{21} &= (M_{t}^{2}M_{t}^{1})(M_{t-\Delta t}^{2}M_{t-\Delta t}^{1}) . . .  (M_{0}^{2}M_{0}^{1}) \\
&= (M_{t}^{2}M_{t}^{1})U_{t}^{21} \\
\Rightarrow dU_{t}^{21} &= (M_{t}^{2}M_{t}^{1})U_{t}^{21} - U_{t}^{21}
\end{split}
\end{equation*}
Expanding the above product using the It\^o table and dropping the superscripts, we arrive at the double-pass QSDE:
\begin{equation}
dU_{t} = \{-L^{*}dA_{t} + LdA_{t}^{*} - \half L^{*}Ldt -iHdt\}U_{t}, 
\end{equation}
with $L = \frac{\alpha(p - ix)}{\sqrt{2}}$, $H = \textstyle{\frac{1}{4}} \alpha^{2}(px + xp)$, and $U_{0} = I$ as before.

\end{document}